\documentclass{PoS}

\PoS{PoS(LAT2005)316}

\usepackage{epsfig}

\title{Study of the systematic errors in the calculation of renormalization constants of the topological susceptibility on the lattice }

\ShortTitle{Systematic errors in lattice renormalization constants
of topological susceptibility}

\author{\speaker{B. All\'es}\\
        INFN Sezione di Pisa, Pisa, Italy\\
        E-mail: \email{alles@df.unipi.it}}
\author{M.~D'Elia\\
        Dipartimento di Fisica, Universit\`a
        di Genova and INFN, Genova, Italy\\
        E-mail: \email{delia@ge.infn.it}}
\author{A. Di Giacomo\\
        Dipartimento di Fisica, Universit\`a di Pisa and
        INFN, Pisa, Italy\\
        E-mail: \email{adriano.digiacomo@df.unipi.it}}
\author{C. Pica\\
        Dipartimento di Fisica, Universit\`a di Pisa and
        INFN, Pisa, Italy\\
        E-mail: \email{pica@df.unipi.it}}

\abstract{We present a study of the systematic effects in the
nonperturbative evaluation of the renormalization
constants which appear in the field--theoretical
determination of the topological susceptibility
in pure Yang--Mills theory. The study is performed by
computing the renormalization constants on configurations
that have been calibrated by use of the
Ginsparg-Wilson formalism and by cooling.}

\FullConference{XXIIIrd International Symposium on Lattice Field Theory\\
25-30 July 2005\\
Trinity College, Dublin, Ireland}

\begin{document}

\section{Introduction}

There are several methods to evaluate the topological susceptibility
$\chi$ in pure Yang--Mills theory on the lattice.
Extracting this quantity for the $SU(3)$ gauge theory is crucial to understand
the large mass of the $\eta'$ meson in QCD~\cite{witten,veneziano}.

Some of the lastest results for $\chi^{1/4}$ obtained on the lattice are:
$177(7)$ MeV by using cooling~\cite{lucini}, $184(7)$ MeV by using
cooling rounding the results to the closest integer~\cite{lucini},
$191(5)$ MeV by counting fermionic zero modes~\cite{deldebbio1} and
$173.4(0.5)(1.2)$ MeV by the field--theoretical (or Pisa) 
method~\cite{alles1} (in this last number the statistical error and the
error derived from the determination of the 
scale $\Lambda_L$ are shown separately).

Any comparison among these methods requires a
good control of all sources of possible systematic errors. We present
a progress report on a study about the systematic errors that are originated
in the evaluation of multiplicative and additive renormalization constants
that appear in the expression of the physical topological susceptibility 
$\chi$ in terms of the lattice one $\chi_L$.
This relationship is~\cite{zimmerman,campostrini2,alles2}
\begin{equation}
\chi_L = Z^2 a^4\chi + M\; ,
\end{equation}
where $Z$ and $M$ are multiplicative and additive renormalization
constants respectively. To extract $\chi$ one has to know the value
of $Z(\beta)$ and $M(\beta)$ where $\beta=6/g^2$, $g$ being the bare
coupling constant. The lattice spacing $a(\beta)$ depends on
the coupling $\beta$ in a way dictated by the lattice beta function.

The lattice susceptibility $\chi_L$ is calculated by using the definition
\begin{equation}
\chi_L\equiv \frac{\left\langle \left(Q_L^{(1)}\right)^2\right\rangle}{L^4}\; ,
\end{equation}
where $L^4$ is the spacetime volume and $Q_L^{(1)}$ is the
1--smeared topological charge~\cite{christou}.

\section{Calculation of the renormalization constants}

The renormalization constants $Z$ and $M$ are evaluated by using
the heating method~\cite{alles3,alles4}. Following
the meaning of $Z$~\cite{campostrini2,alles2}, we compute
the average topological charge within a fixed topological sector.
If we choose a topological sector of charge $n$ (any nonzero integer) then
\begin{equation}
Z=\frac{\left\langle Q_L^{(1)}\right\rangle\Big\vert_{Q=n}}{n} \; ,
\end{equation}
where the division by $n$ entails the requirement that $Q$ takes
integer values ($Q$ here is defined by cooling). The brackets
$\langle\cdot\rangle |_{Q=n}$
mean thermalization within the topological sector of charge $n$.

We start our algorithm with a classical configuration with
topological charge $1$ ($Q=1$) and action $8\pi^2$ in 
appropriate units.\footnote{The numerical values of $Q$ and the
action have been calculated by using $Q_L^{(1)}$ and the Wilson pure gauge
action~\cite{wilson} on a cooled 1--instanton configuration. The
actual numbers turn out to be not exactly $1$ and $8\pi^2$ but very
close to them. The differences vanish in the thermodynamic limit.}
Then we apply $80$ heat--bath
steps of Cabibbo--Marinari updating~\cite{cabibbo} 
and measure $Q_L^{(1)}$ every $4$ steps. 
This set of $20$ measurements is called ``trajectory''.
After each measurement we apply 8 cooling steps
to verify that the topological sector is not changed. We repeat
this procedure to obtain a number of trajectories. For each trajectory we
always discard the first few measurements because the configuration is not
yet thermalized. Averaging over the thermalized steps
(as long as the corresponding cooled configuration shows the correct
background topological charge, $Q=1$ within a deviation $\delta$)
yields $\left\langle Q_L^{(1)}\right\rangle\Big\vert_{Q=1}$. We estimate
the systematic error that stems from the choice 
of $\delta$ as in~\cite{alles4}.

As for the additive renormalization constant $M$ the procedure is
quite analogous. We calculate 
$M=\chi_L|_{Q=0}\equiv\left\langle\left(Q_L^{(1)}\right)^2
\right\rangle\Bigg\vert_{Q=0}/L^4$
in the zero topological charge sector (because within this sector
the physical topological susceptibility $\chi$ vanishes).
Single trajectories consist again of $80$ heat--bath steps with measurements
every $4$ steps and cooling tests after each measurement. Thermalization
of short distance fluctuations again requires discarding a few initial
steps.

When a cooling test reveals that the corresponding configuration
lies outside the correct topological sector ($Q=0$ for the
calculation of $M$ and $Q=1$ for $Z$) then all points of the
trajectory are discarded from that configuration onwards. In full
QCD this event seldom happens~\cite{vv1,vv2}; we work however in
pure Yang--Mills theory where it happens more frequently.

\begin{figure}
\vspace{0.7cm}
\hspace{2cm}
\epsfig{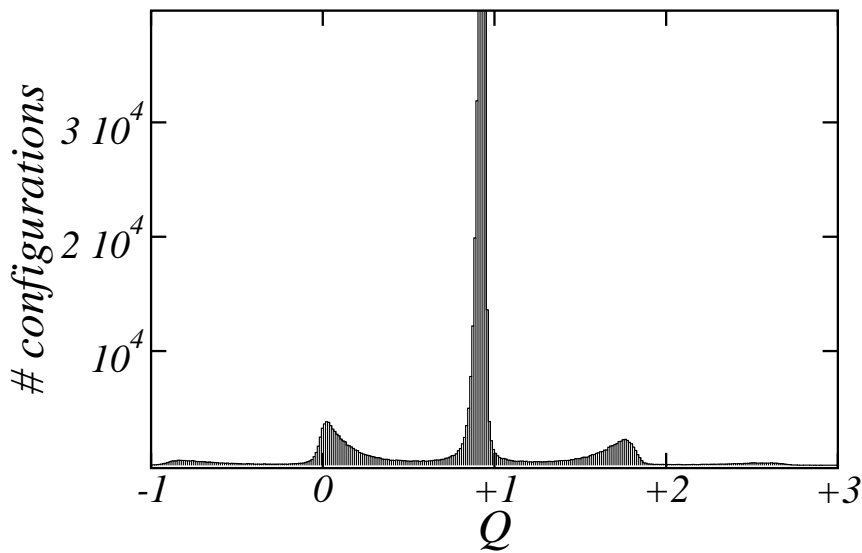}
\caption{Histogram of background topological charge on all
configurations obtained during a calculation of $Z(\beta=6.0)$
on a $16^4$ lattice.}
\label{fig1}
\end{figure}

An histogram containing the results of all cooling tests during 
a calculation of $Z(\beta=6.0)$
is shown in Fig.~\ref{fig1}. 
It has been drawn from the set of all configurations measured within
$25000$ trajectories: $25000\times 20=5\;10^5$ configurations. It is
clear that in some cases the topological sector has moved towards
topological charges different from $+1$. This occurs in
the $30\%$ of all cases.
Notice that the histogram is slanted: there are more configurations
moving towards $Q=0$ than towards $Q=+2$ because the system tends to
equilibrate around $Q=0$. Therefore if configurations within the
wrong topological sector were not discarded, we would obtain a biased
value for $Z$ (lower than the correct one).

When the additive renormalization $M$ is calculated, all configurations
should lie in the $Q=0$ sector. If configurations that migrated outside 
this sector were not eliminated, the outcome for $M$ would be larger than the
correct one as any departure from zero increases $M$ because it
is proportional to the square of $Q_L^{(1)}$.

However the use of cooling to unravel the topological sector of
a configuration may also destroy the unwanted instantons that
alter the correct sector. In such a case we could not be able to
discover that we are calculating the renormalization constants
on wrong topological backgrounds. This problem comes about when
the unwanted instanton is small and few cooling steps can destroy
it rather easily. As described above, this
circumstance distorts the extraction of $Z$ and $M$,
yielding values for these constants that are lower ($Z$) or larger
($M$) than the correct ones. This is the source of systematic error
that we want to quantify by checking the background topological
charge by using a method different from cooling.

The counting of fermionic zero modes is a method to perform this check
without modifying the configuration. Unfortunately it is 
much more demanding than cooling for quantifying the above described
systematic effect.

\section{Zero modes counting checks}

The net number of zero modes $n_+ - n_-$ is obtained by counting
the level crossings in the spectrum of the Wilson--Dirac operator
as the fermion mass $m$ varies~\cite{narayanan,neuberger,edwards}.
This method was used in~\cite{edwards,deldebbio2} to calculate
the topological charge. We use an accelerated conjugate gradient
algorithm~\cite{kalkreuter} to extract the lowest eigenvalues of the
Wilson--Dirac operator.

This technique looks ambiguous because there can be level crossings
all along the interval of masses where the gap is closed. If we
stop counting crossings at some mass inside the interval, 
in general the resulting topological charge depends 
on the choice of this mass. It is shown~\cite{edwards}
that physical results do not depend strongly on the choice of this
mass. In particular, instantons representing crossings that are
close to $a\,m=2$ (the endpoint of allowed masses) have a size
of a few lattice spacings. These are precisely the
instantons that most probably could evade the cooling test. Therefore we
show results for $a\,m=2$.

\begin{figure}
\vspace{0.7cm}
\hspace{2cm}
\epsfig{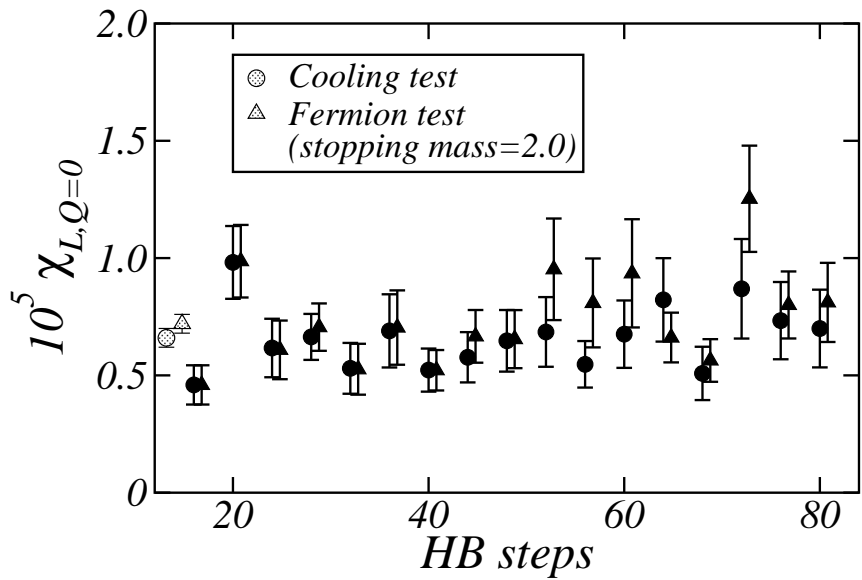}
\caption{Average on a set of $50$ trajectories for the calculation
of $M$ by using both cooling (circles) and fermionic level crossing counting
(triangles) as checks of the topological background. The level cross
counting stopped at $a\,m=2$.}
\label{fig2}
\end{figure}

We want to calculate both $M$ and $Z$ at some fixed value of the
lattice coupling $\beta$. The topological charge
background will be checked both by cooling and by counting of
fermionic zero modes and the final results compared. Any difference
between these final results can be interpreted as an estimate
of the systematic error.
At this writing we have
completed $50$ trajectories for $M$ at $\beta=6.0$ on a $12^4$
lattice for the quenched $SU(3)$ theory.
We show the partial results deduced
from this limited set of trajectories. In Fig.~\ref{fig2} 
the average among trajectories
is shown for the two methods of checking.
As for the cooling, the configuration
was considered as belonging to the right sector $Q=0$ 
within a deviation $|\delta | =0.3$.
We show only data after the $16$th heat--bath step (previous steps
are irrelevant because the quantum fluctuations are not yet thermalized).
The points in grey color are the average of all data in all trajectories.
They are
\begin{equation}
M_{\rm cooling}=0.66(4) \; 10^{-5}\; , \qquad\qquad
M_{\rm fermionic}=0.72(4) \; 10^{-5}\; .
\label{Mcool}
\end{equation}

\begin{figure}
\vspace{0.7cm}
\hspace{2cm}
\epsfig{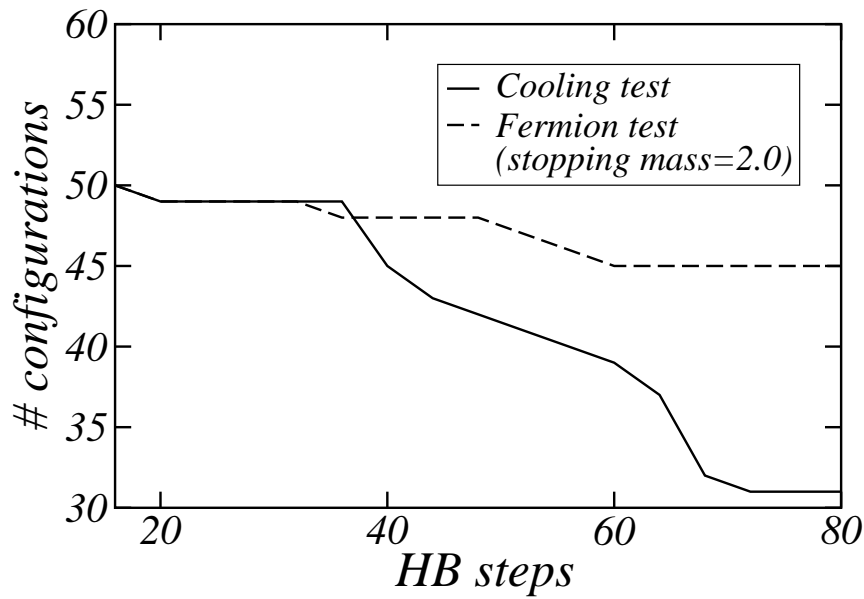}
\caption{Number of configurations used for the average in
Fig.~3 as a function of the heat--bath
step for cooling checks (straight
line) and fermionic level crossing counting (dashed line).}
\label{fig4}
\end{figure}

In Fig.~\ref{fig4} the number of configurations
used for the average in Fig.~\ref{fig2}
at each step is displayed. Ideally the number of configurations should be
$50$, however due to the discarding of configurations with a wrong
topological charge background, this number decreases as a function
of the step. Notice that this decrease is steeper
for the cooling although it is supposed that some instantons
escape the cooling check while they should be netted by the
counting of zero modes (mainly with the stopping mass fixed at $a\, m=2$).
The plot depends little on the choice of $\delta$, ($\delta=0.1$ or $0.5$
yield similar data).
The fact that the result for $M$ obtained
with cooling (see~(\ref{Mcool})) tends to be lower than the
results obtained by counting zero modes seems to indicate a better
performance of cooling in discovering wrong charge configurations.
However our error bars are still too large to be able to draw any clean
conclusions from the data. We plan to improve the statistics roughly by 
one order of magnitude both for $M$ and $Z$.


\end{document}